\documentclass[conference]{IEEEtran}
\IEEEoverridecommandlockouts
\usepackage{cite}
\usepackage{amsmath,amssymb,amsfonts,hyperref}
\usepackage{algorithmic}
\usepackage{graphicx}
\usepackage{textcomp}
\usepackage{xcolor}
\usepackage{xspace}
\usepackage{float}
\usepackage{caption}
\usepackage{subcaption}
\usepackage{makecell}
\usepackage{booktabs}
\usepackage{xcolor}
\usepackage{balance}
\usepackage{tcolorbox}
\usepackage{listings}
\usepackage[inline]{enumitem}

\definecolor{codegreen}{rgb}{0,0.6,0}
\definecolor{codegray}{rgb}{0.5,0.5,0.5}
\definecolor{codepurple}{rgb}{0.58,0,0.82}
\definecolor{backcolour}{rgb}{245,245,245}

\lstdefinelanguage{json}{
    basicstyle=\ttfamily\scriptsize,
    numbers=left,
    numberstyle=\tiny\color{codegray},
    stepnumber=1,
    numbersep=5pt,
    showstringspaces=false,
    breaklines=true,
    frame=single,
    backgroundcolor=\color{backcolour},
    literate=
     *{:}{{{\color{red}{:}}}}{1}
      {,}{{{\color{red}{,}}}}{1}
      {\{}{{{\color{red}{\{}}}}{1}
      {\}}{{{\color{red}{\}}}}}{1}
      {[}{{{\color{red}{[}}}}{1}
      {]}{{{\color{red}{]}}}}{1},
}

\lstdefinestyle{mystyle}{
    backgroundcolor=\color{backcolour},   
    commentstyle=\color{codegreen},
    keywordstyle=\color{magenta},
    numberstyle=\tiny\color{codegray},
    stringstyle=\color{codepurple},
    basicstyle=\ttfamily\scriptsize,
    breakatwhitespace=false,         
    breaklines=true,                 
    captionpos=b,                    
    keepspaces=true,                 
    numbers=left,                    
    numbersep=5pt,                  
    showspaces=false,                
    showstringspaces=false,
    showtabs=false,                  
    tabsize=4
}

\newcommand{\benchmarkname}{{\sc BUMP}\xspace}
\newcommand{\nbTotalReproductions}{5364\xspace}
\newcommand{\nbBreakingUpdates}{571\xspace}
\newcommand{\nbBreakingUpdatesBeforeSanityCheckings}{628\xspace}
\newcommand{\nbUniqueProjects}{153\xspace}

\newcommand{\nbBreakingUpdateImages}{1142\xspace}
\newcommand{\nbDirectDependencies}{16}
\newcommand{\nbTransitiveDependencies}{68}

\newcommand{\nbMedianDockerImages}{733.15\xspace}

\newcommand{\biapi}{\texttt{biapi}\xspace}

\newcommand{\etal}{et al.\@\xspace}

\newcommand{\nbDependencyChangesPerBDU}{2\xspace}

\newcommand{\pom}{\texttt{pom.xml}\xspace}
\newcommand{\precomm}{pre-breaking commit\xspace}
\newcommand{\breakingcomm}{breaking-commit\xspace}

\newcommand{\nbCannotFindSymbol}{1950}
\newcommand{\nbCannotAccess}{419}
\newcommand{\nbPackageNotExist}{344}
\newcommand{\nbConstructor}{129}
\newcommand{\nbMethOverride}{39}
\newcommand{\nbStaticImport}{16}
\newcommand{\nbIncompatibleTypes}{13}
\newcommand{\nbReferencesAmbiguous}{12}
\newcommand{\nbMethodTypes}{9}

\newcommand{\nbCompilationBreakage}{243\xspace}
\newcommand{\nbTestingBreakage}{188\xspace}
\newcommand{\nbMavenEnforcerBreakage}{121\xspace}
\newcommand{\nbDependencyResolutionBreakage}{5\xspace}
\newcommand{\nbDependencyLockBreakage}{14\xspace}

\newcommand{\percentageCompilationTestBreakage}{76}
\newcommand{\percentageEnforcerBreakage}{21}
\newcommand{\percentageCompilationBreakage}{43}
\newcommand{\percentageTestBreakage}{33}
\newcommand{\percentageAllOtherBreakage}{24}
\newcommand{\percentageDependencyLockFailure}{2}
\newcommand{\percentageDependencyResolutionFailure}{1}

\def\commits{100}

\newcommand\bumpurl{\href{https://github.com/chains-project/bump}{https://github.com/chains-project/bump}\xspace}

\newtheorem{definition}{Definition}

\def\BibTeX{{\rm B\kern-.05em{\sc i\kern-.025em b}\kern-.08em
    T\kern-.1667em\lower.7ex\hbox{E}\kern-.125emX}}
\begin{document}
\setlength{\abovecaptionskip}{1mm plus 0mm minus 0mm}
\makeatletter
\newcommand{\linebreakand}{%
  \end{@IEEEauthorhalign}
  \hfill\mbox{}\par
  \mbox{}\hfill\begin{@IEEEauthorhalign}
}
\makeatother

\title{\benchmarkname: A Benchmark of Reproducible Breaking Dependency Updates}

\author{\IEEEauthorblockN{1\textsuperscript{st} Frank Reyes}
\IEEEauthorblockA{\textit{KTH Royal Institute of Technology}\\
Stockholm, Sweden \\
frankrg@kth.se}
\and
\IEEEauthorblockN{2\textsuperscript{nd} Yogya Gamage}
\IEEEauthorblockA{\textit{KTH Royal Institute of Technology}\\
Stockholm, Sweden \\
yogya@kth.se}
\and
\IEEEauthorblockN{3\textsuperscript{rd} Gabriel Skoglund}
\IEEEauthorblockA{\textit{KTH Royal Institute of Technology}\\
Stockholm, Sweden \\
gabsko@kth.se}
\linebreakand
\IEEEauthorblockN{4\textsuperscript{th} Benoit Baudry}
\IEEEauthorblockA{\textit{KTH Royal Institute of Technology}\\
Stockholm, Sweden \\
baudry@kth.se}
\and
\IEEEauthorblockN{5\textsuperscript{th} Martin Monperrus}
\IEEEauthorblockA{\textit{KTH Royal Institute of Technology}\\
Stockholm, Sweden \\
monperrus@kth.se}

}
\maketitle

\begin{abstract}

Third-party dependency updates can cause a build to fail if the new dependency version introduces a change that is incompatible with the usage: this is called a breaking dependency update.
Research on breaking dependency updates is active, with works on characterization, understanding, automatic repair of breaking updates, and other software engineering aspects. All such research projects require a benchmark of breaking updates that has the following properties:
1) it contains real-world breaking updates;
2) the breaking updates can be executed;
3) the benchmark provides stable scientific artifacts of breaking updates over time, a property we call ``reproducibility''.
To the best of our knowledge, such a benchmark is missing.
To address this problem, we present \benchmarkname, a new benchmark that contains reproducible breaking dependency updates in the context of Java projects built with the Maven build system.
\benchmarkname contains \nbBreakingUpdates breaking dependency updates collected from \nbUniqueProjects Java projects.
\benchmarkname ensures long-term reproducibility of dependency updates on different platforms, guaranteeing consistent build failures. We categorize the different causes of build breakage in \benchmarkname, providing novel insights for future work on breaking update engineering.
To our knowledge, \benchmarkname is the first of its kind, providing hundreds of real-world breaking updates that have all been made reproducible.
\end{abstract}


\section{Introduction}
In software development, software projects increasingly rely on external dependencies, leveraging code reusability \cite{soto2023automatic}
Hence, dependency management is a critical aspect of ensuring the stability \cite{pashchenko/2020}, security \cite{chinthanet/2021}, and durability \cite{digkas2018developers} of software.
Keeping dependencies up-to-date is essential in order to fully benefit from code reuse \cite{kula2018developers}. However, the process of updating outdated dependencies is not as straightforward as merely changing the version number \cite{javan2023dependency, he2023automating}. This complexity results from the potential risk that the updated version may introduce breakages to the project. Concretely, these breakages might occur because of two main reasons: changes in the dependency interface (API) and, changes in the behavior of the updated dependencies \cite{ganea2017hindsight}: changes could be syntactic, such as adding or deleting a class, or behavioral, such as affecting the side effects of a method.

Several studies have analyzed the problem of breaking dependency updates on software development \cite{ochoa2022breaking,mirhosseini2017can,soto2021longitudinal}. A major challenge arises when attempting to compare these findings, as each study has utilized its own dataset. This poses a threat to the scientific validity of these results.
Moreover, we note that it is impossible to reproduce some of them, as the study subjects depended on several uncontrolled environmental factors such as the state of the package manager at one point in time and the state of the operating system being used.
This poses a threat to reproducible science. 
To mitigate these threats, there is an urgent need for a benchmark that comprises real-world, reproducible breaking updates.
Such a benchmark would ensure the long-term reproducibility of any research made in the important field of software dependency engineering research. 
 
In this paper, we introduce a benchmark of breaking dependency updates in real-world Java projects, referred to as \benchmarkname. It consists of a collection of \nbBreakingUpdates breaking dependency updates collected from \nbUniqueProjects Java projects that build with Maven.
Each breaking update in \benchmarkname is stored within Docker images, ensuring long-term reproducibility. \benchmarkname enables users to reproduce the passing build before the dependency update and the breaking build after the dependency update by running the two Docker images as containers.

To ensure that the \benchmarkname can be successfully utilized for future research, we run extensive experiments on the reproducibility of the breaking updates in \benchmarkname. 
The obtained results confirm that the breaking updates can be fully reproduced, both on Linux and Windows.
Also, we demonstrate that network connection can be cut-off, demonstrating that the Docker images completely capture the artifacts needed to build the project.
We perform a categorization of the failures caused by breaking dependency updates, and our results report that 76\% of the failures are caused by compilation and test errors. The other 24\% of failures contain enforcer rule violations and dependency locking failures that were never studied in previous research.

To the best of our knowledge, \benchmarkname is the first ever dataset that consists of real-world breakages from dependency updates, with the guarantee of long-term reproducibility. Furthermore, we present original findings on breaking update failure types and the underlying causes of build failures that allow researchers to better understand the nature of breaking updates.

To summarize, our key contributions are:
\begin{itemize}
    \item An original methodology for collecting reproducible breaking dependency updates, with special care taken in addressing the main threats to reproducibility: package manager transient state, flaky tests, and operating system dependence.
    
    \item A benchmark called \benchmarkname: it contains \nbBreakingUpdates reproducible breaking dependency updates collected from \nbUniqueProjects Java projects built with Maven.

    \item Guarantees of reproducibility: all breaking updates are put in Docker images that can be run offline, providing long-term replicability, available at \bumpurl.

    \item A systematic characterization of the symptoms and causes of breaking dependency updates: we are the first to report on the existence of dependency update build failures due to dependency rules and dependency locking practices.
    
\end{itemize}

\begin{figure*}
\centering
    \begin{subfigure}{0.33\textwidth}
    \includegraphics[width=1\linewidth]{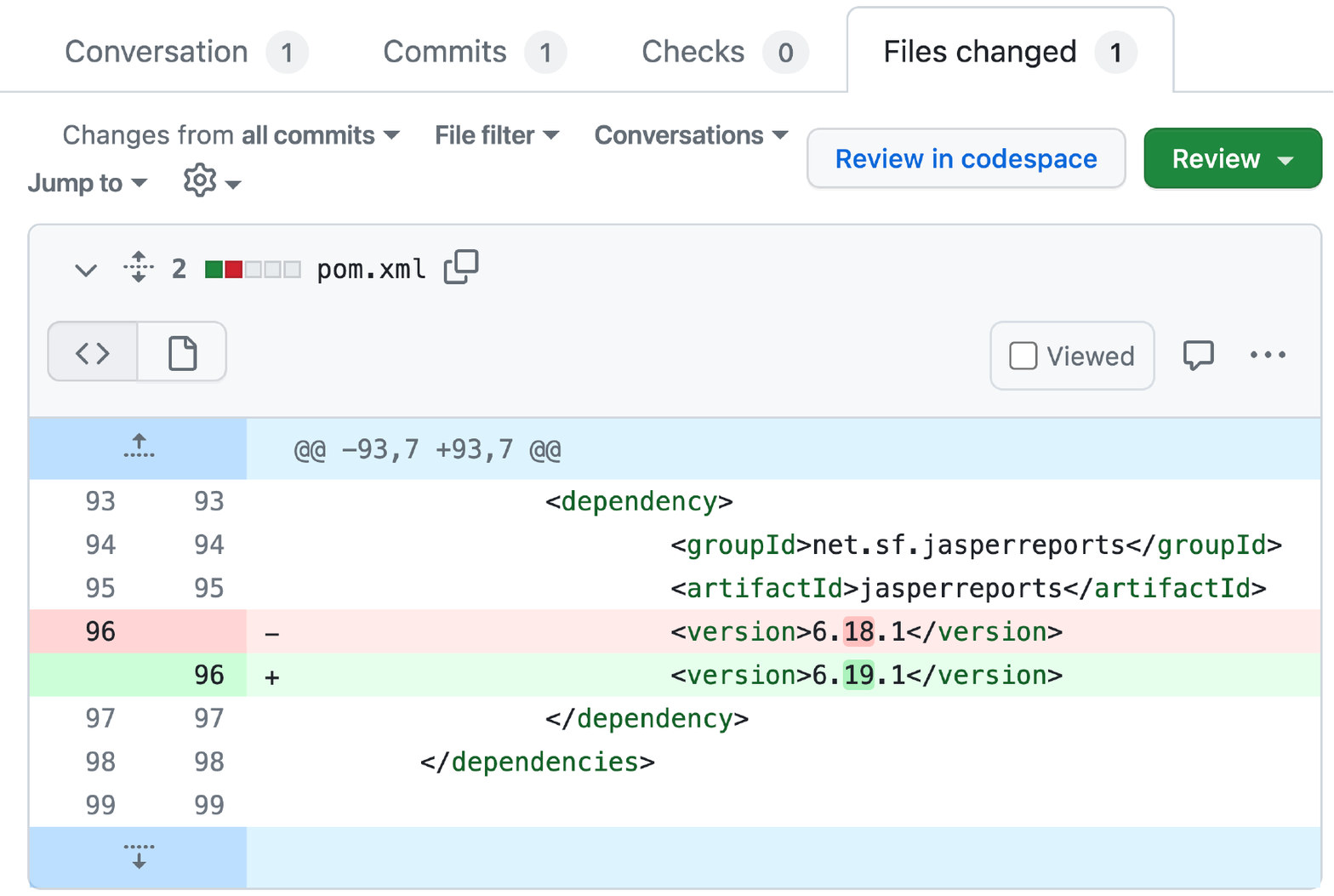}
    \caption{Dependency version update in Maven build file}
    \label{fig: updated}
    \end{subfigure}
    \hfill
    \hspace{-1cm}
    \begin{subfigure}{0.33\textwidth}
    \includegraphics[width=1\linewidth]{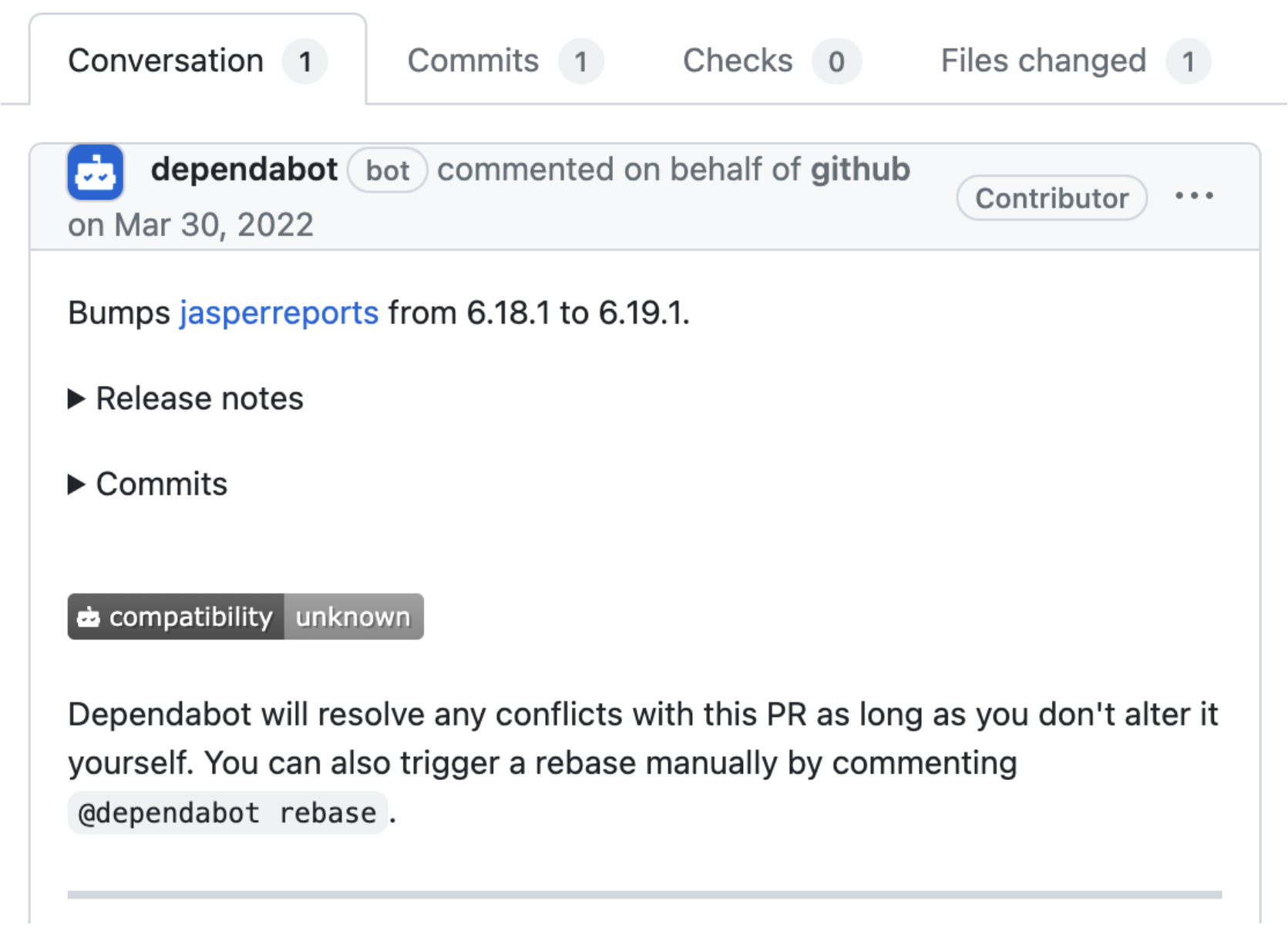}
    \caption{Information provided by the bot that suggested the dependency update}
    \label{fig: conversation}
    \end{subfigure}
    \hfill
    \hspace{-1cm}
    \begin{subfigure}{0.33\textwidth}
    \includegraphics[width=1\linewidth]{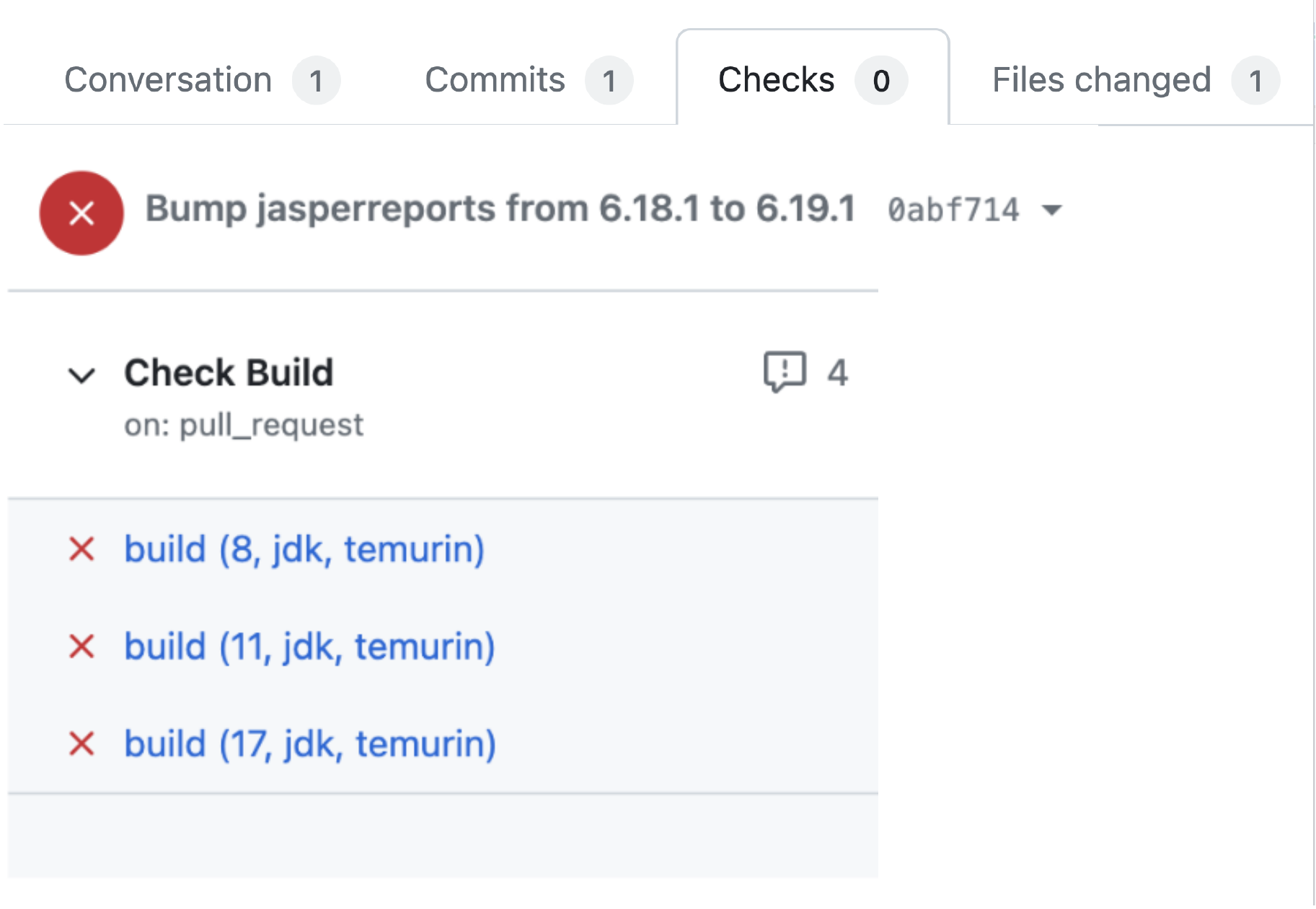}
    \caption{GitHub action failure after the dependency update}
    \label{fig: check}
    \end{subfigure}
    \caption{A real-world example of a breaking update.}
    \label{fig:pullrequest}
\end{figure*}

\begin{figure*}
\centering
\includegraphics[width=1\linewidth]{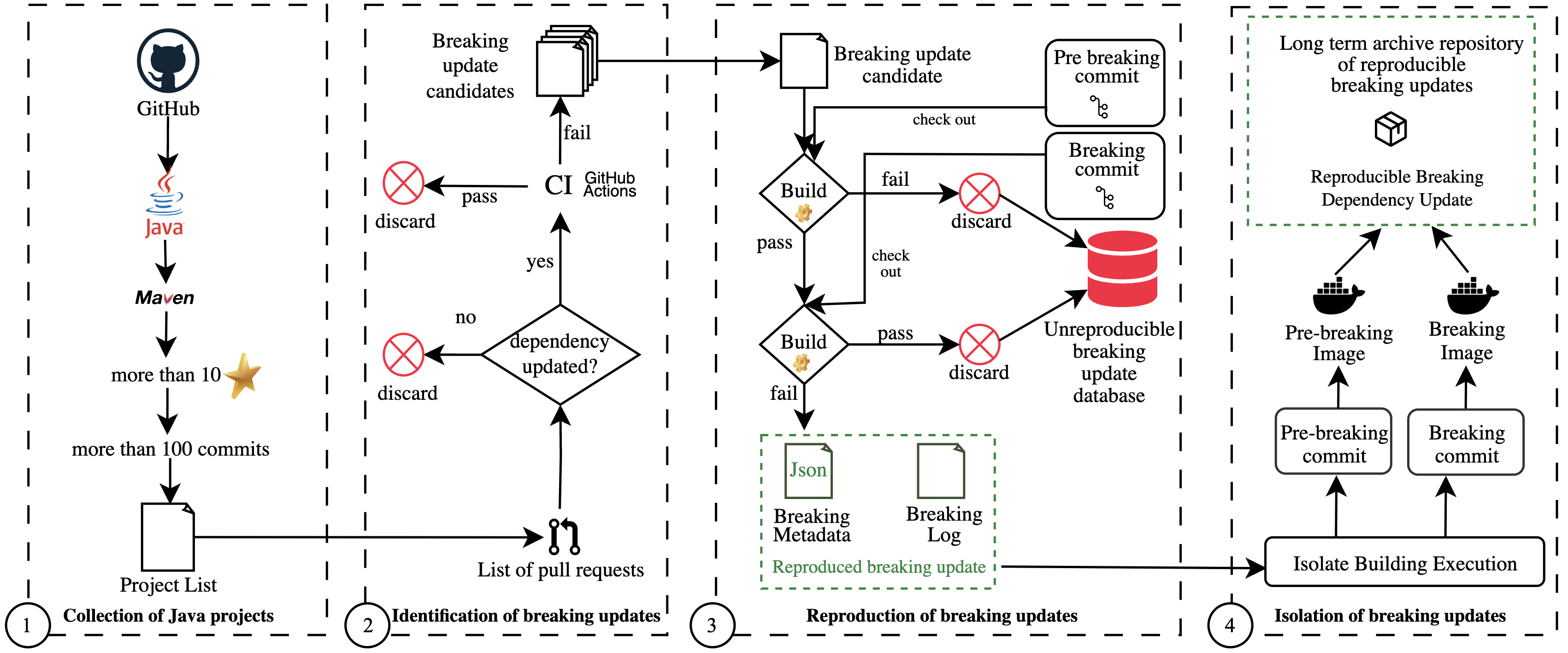}
\caption{Overview of the methodology to build the \benchmarkname benchmark.}
\label{fig:methodology}
\end{figure*}

\section{Methodology for Buidling the \benchmarkname Benchmark}
\label{methodology}

In this section, we introduce the steps we propose to build a benchmark of breaking dependency updates, as well as the main design decisions to make the benchmark relevant and reproducible.

\subsection{Concepts}

We aim to collect a set of breaking dependency updates and consolidate this set such that all breaking updates are fully reproducible.
A breaking update results from the publication of a non-backward compatible library version, called hereafter a breaking dependency version.

\begin{definition}
\label{brekDepversion}
\textbf{A breaking dependency version} is a library version in a package manager with a breaking API or a breaking behavioral change.
\end{definition}

Clients typically update a build configuration file to bump dependency versions. In \autoref{fig:pullrequest}, we provide an example of a \autoref{brekDepversion} found in the \href{https://github.com/xdev-software/biapi}{\biapi} project hosted on GitHub. \autoref{fig: updated} illustrates the modification of the dependency version number in the project's build configuration file, meanwhile, \autoref{fig: conversation} contains the information provided by the author of the update. \autoref{fig: check} illustrates the build failure for the dependency update in the \biapi project.

In this context, we designate a \breakingcomm as a commit that only modifies the version of an existing dependency in the build configuration file. The \precomm is a commit that precedes the \breakingcomm. When we build the project at the \precomm, we expect the compilation and testing execution to be successful.

That is, we define a breaking dependency update as:
\begin{definition}
    \label{breakdepupdate}
    A \textbf{breaking dependency update} is a pair of commits for a project composed by a \precomm with a passing build and a \breakingcomm with a failing build, such that the version of one single dependency is updated.
\end{definition}

\autoref{fig:methodology} shows the key steps to build our novel benchmark of breaking dependency updates.
The process to build \benchmarkname starts with collecting Java projects on GitHub (step 1). For each project, we look at its history for a \breakingcomm (step 2). If we find a \breakingcomm, we proceed to look for the \precomm and we try to build it (step 3). If we succeed in building it, we store the breaking update in a container, which supports portability as well as reproducibility (step 4). 
Each step is discussed in detail in the following sections.

\subsection{Collection of Java projects}

We collect Java projects that are built with Maven, and that meet the following criteria: at least \commits\xspace commits on the default branch, created in the last 10 years, at least 3 contributors, at least 10 stars. These criteria are meant to filter out toy projects for which breaking dependency updates are irrelevant. 

We pass the set of GitHub projects that fulfil these criteria to step 2, where we search for breaking updates. In the following subsections, we illustrate each step of our process with the \texttt{biapi} project. 
As of March 14, 2023, this project has 178 commits, 7 contributors, and 19 stars. \texttt{biapi} meets the necessary criteria to be analyzed by \benchmarkname. As of Aug 10 2023, it contains 18 candidate pull requests for breaking dependency updates.

\subsection{Mining of breaking update candidates}

We analyze the set of pull requests of each Java project in order to find breaking dependency updates. A pull request is a candidate breaking update if it satisfies the following criteria: 1) the pull request only modifies the \pom file, 2) the modification is a one-liner, 3) it changes the version number of a dependency and 4) the pull request fails to pass the build.
In this context, we define a breaking dependency candidate as: 

\begin{definition}
\label{BDU}
\textbf{A breaking update candidate} is a pull request in a Java project, that only contains one \breakingcomm. When we build the project with the \breakingcomm, compilation or test execution fails. 
\end{definition}

For example, pull request \href{https://github.com/xdev-software/biapi/pull/69}{\#69} of \texttt{biapi} satisfies all criteria. The pull request is illustrated in \autoref{fig:pullrequest}. It includes one commit, which modifies the \pom file to update the version of \texttt{jasperreports} from 6.18.1 to 6.19.1 and fails to build.

We discard any pull request that updates a test scope dependency, as they are only included in the project classpath during the test phases and are irrelevant when deploying the project. Step 2 results in a set of Java projects which include at least one pull request that is a potential breaking dependency update.

\subsection{Reproduction of breaking updates}
\label{sec:reproduction-process}

The third step of \benchmarkname is the reproduction of breaking updates.
At this point, we aim to recreate, in a local environment, the conditions to reproduce a breakage on the same dependency update commit which triggered the failure in CI. We define this process as a reproduction of a breaking dependency update.

\begin{definition}
\label{reproduceBDU}
\textbf{A reproduced breaking dependency update} is a breaking dependency update that meets the following constraints:
1) the \precomm has passed in a local environment;
2) the \breakingcomm has failed in a local environment, and the failure is causally related to the dependency update.

Here, a local environment means a Docker container with the cloned GitHub repository and the configuration to build a Maven project using Java version 11\footnote{Docker base image is specified in \autoref{isolation} }. 
\end{definition}

Given a Java project and a candidate pull request, we first identify the commit that is immediately preceding the last commit of the pull request, which we refer to as \precomm. Then, we try to build the project on that commit.
If the build gets successful, then we consider this commit as a potential \precomm.

Second, we try to build the project with the commit that breaks the build on CI. At this stage, the compilation can fail, or at least one test case can fail. We set a 10-hour timeout for the whole build process. 

If we succeed in having a \precomm that systematically passes the build and a \breakingcomm that systematically fails the build, we consider this pair of commits as a reproduced breaking dependency update as in \autoref{reproduceBDU}. 

In the \texttt{biapi} project we take as an example in \autoref{fig:pullrequest}, the \precomm is
\href{https://github.com/xdev-software/biapi/commit/b508900064878eb1ed55a53a724da2fe58f56ad0}{
b508900064878eb1ed55a53a724da2fe58f56ad0} and the \breakingcomm is \href{https://github.com/xdev-software/biapi/pull/69/commits/0abf7148300f40a1da0538ab060552bca4a2f1d8}{0abf7148300f40a1da0538ab060552bca4a2f1d8}, and the breakage is a compilation error \emph{incompatible types: int cannot be converted to java.lang.Float}.

Some \precomm builds fail in our local environment. This is because it is not always feasible to replicate all the conditions under which the project is built within CI. For example, some projects require access to external resources such as databases or authorized web APIs to test the project. In all cases, we consider the breaking dependency update as an unreproducible one if we do not manage to build the \precomm. We store the data to analyze the causes of reproducibility.

The opposite problem is when the \breakingcomm passes the local build while failing in CI. This happens because the version of OpenJDK where the project is compiled within CI is not the same as in the local environment where the pull request is reproduced. We met this anomaly in the \href{https://github.com/awhitford/lombok.maven/pull/178}{lombok.maven} project. The compilation process within the CI utilizes Java 8 and \benchmarkname employs version 11 in the local environment. As the build concludes successfully even with the \breakingcomm for those candidates, we discard them. 
After step 3, we have a list of successfully reproduced breaking dependency updates.

\subsection{Isolation and long-term archival of dependencies}
\label{isolation}

The final step for the construction of \benchmarkname consists of isolating each reproduced breaking update. With this step, we want to ensure the durability and reproducibility of our benchmark. We rely on Docker for isolation. We build a pair of Docker images: one to reproduce the \precomm and another one for the \breakingcomm. In light of these objectives, we define this process as:

\begin{definition}
\label{reproducibleBDU}
\textbf{A reproducible breaking dependency update} is a breaking dependency update that has been successfully reproduced and that also meets the following constraints:
1) all dependencies have been extracted and saved for long-term preservation.
2) the complete building pipeline (compiler, build toolchain) has been saved on disk and long-term preserved.
In the \benchmarkname benchmark, a reproducible breaking dependency update is stored as a pair of Docker images that can reproduce the \precomm and the \breakingcomm.
\end{definition}

We build the Docker images with respect to two properties: size and reproducibility. The images should be compact, which is essential for optimizing storage efficiency, expediting distribution, and minimizing resource utilization during deployment. The images should be capable of successfully reproducing Java breaking updates. This requires the inclusion of fundamental dependencies such as OpenJDK, Git, and Apache Maven. 

In order to meet these requirements, we first create a base image, using a vanilla Alpine Linux image (7.33MB). Then, we install OpenJDK version 11.0.19, Git version 2.40.1, and Apache Maven 3.9.2. At this point, the base image is 292MB. Following the base image creation, we run it as a container and clone the GitHub project we want to build. We use shallow cloning with a depth of 2 to avoid image sizes getting excessively large \cite{shadow}.
Then, we checkout the \breakingcomm. To compile and test the \precomm, we move back one commit to the immediate parent commit using the \texttt{git checkout HEAD~1} command. Subsequently, we run the \texttt{mvn clean test -B} command and save the log result. We follow the same procedure for the \breakingcomm in another container, excluding the step to move the HEAD back one commit. To further reduce the size of the images, we delete the .git folder afterward. Finally, we create two new images with the entry command \texttt{mvn clean test -B} from the two containers for reproduction in the future. 

Each Docker image is named according to the following convention: id of the commit that breaks the build, followed by \emph{-pre} or \emph{-breaking}. 
Docker's portability enables consistent project execution across diverse operating systems and configurations.
The created images to reproduce the build for the \precomm and the \breakingcomm of the example project \texttt{biapi} are available \href{https://github.com/orgs/chains-project/packages/container/breaking-updates/123314023?tag=0abf7148300f40a1da0538ab060552bca4a2f1d8-pre}{online}.
All the created images, including the base image are stored on the GitHub Docker registry, offering a secure and scalable storage space. 
They are also pushed on Zenodo for long-term preservation in case the GitHub Docker registry disappears.

\subsection{Sanity checks}
\label{sec:sanity}

To ensure the fundamental properties of the benchmark, we perform the following sanity checks. 
Flaky tests are a known problem in real-world builds. They might affect both pre-breaking images and breaking ones. For the former, the pre-breaking image might fail while the project does not actually contain errors.
In the latter case, we might observe failures in the breaking image that are not causally related to the breaking update.
To overcome this problem, we run the test suite three times on the pre-breaking and the breaking images to ensure a consistent absence of failures, resp. causal failures.
Second, the other major hindrance to reproducibility is the dependency on the operating system. For instance, one project might break on Linux but pass on Windows. To overcome this, we run all pre-breaking and breaking images on both platforms three times.
Third, we note that some breakages are not due to the dependency update per se, but are caused by the build system configuration that is not exactly the right one. In particular, this happens with build configurations related to the Jaxb annotation framework. We manually discard those projects during post-processing.
We also observe a long tail of rare and exotic failures, such as the ones related to \texttt{org.apache.maven.scm/maven-scm-provider-jgit} which are not related to application dependencies. Therefore, they are also discarded.

\subsection{Data format}
\label{sec:data-format}

The \benchmarkname benchmark is available as a collection of metadata files, where each file is named according to the unique hash corresponding to the \breakingcomm. For example, in \autoref{lst:example-json-file} we show the metadata for the illustrative breaking update in \texttt{biapi}:\emph{0abf7148300f40a1da0538ab060552bca4a2f1d8.json}.

The \emph{prAuthor}, \emph{preCommitAuthor}, and \emph{breakingCommitAuthor} define whether the pull request, previous commit, and the breaking commit are authored by a human or a bot such as Renovate or dependabot. We retrieve this author-type information by querying the user type from the GitHub API for the corresponding author, and by checking whether the username of the author contains the names \texttt{renovate} or \texttt{dependabot}. 

Each Maven artifact is uniquely identified as a tuple consisting of a groupId, an artifactId and a version number. We extract these values to document the \emph{updatedDependency} as its \emph{dependencyGroupID}, \emph{dependencyArtifactID}, \emph{previousVersion} and the \emph{newVersion} of the dependency that has been updated as part of the \breakingcomm. For example, the breaking update in \texttt{biapi}, changes the version of \texttt{net.sf.jasperreports.jasperreports} from the \emph{previousVersion} \texttt{6.18.1} to the \emph{newVersion} \texttt{6.19.1}. If the \emph{previousVersion} and the \emph{newVersion} follow the pattern \emph{Major.Minor.Patch} we can determine the \emph{versionUpdateType}. If the versions follow the pattern \emph{Major.Minor}, we follow the same procedure assuming the patch version number is 0. For the versions that do not follow either of these patterns, \emph{versionUpdateType} is considered as \emph{other}. The \emph{dependencyScope} is the Maven scope of the updated dependency. If the scope is not defined we \emph{compile} as the default scope. The \emph{updatedFileType} is the type of the updated dependency: a pom type dependency or a default jar type dependency. The \emph{dependencySection} records the section under which the dependency is declared in the \pom file such as the dependency management section or build section. 

\begin{lstlisting}[language=json, caption={Scientific metadata for a reproduced breaking update in project biapi}, label=lst:example-json-file]
{
  "url": "https://github.com/xdev-software/biapi/pull/69",
  "project": "biapi",
  "projectOrganisation": "xdev-software",
  "breakingCommit": "0abf7148300f40a1da0538ab060552bca4a2f1d8",
  "prAuthor": "bot",
  "preCommitAuthor": "human",
  "breakingCommitAuthor": "bot",
  "updatedDependency": {
    "dependencyGroupID": "net.sf.jasperreports",
    "dependencyArtifactID": "jasperreports",
    "previousVersion": "6.18.1",
    "newVersion": "6.19.1",
    "dependencyScope": "compile",
    "versionUpdateType": "minor",
    ...
    "updatedFileType": "JAR"
  },
  "preCommitReproductionCommand": "docker run ghcr.io/chains-project/breaking-updates:0abf7148300f40a1da0538ab060552bca4a2f1d8-pre",
  "breakingUpdateReproductionCommand": "docker run ghcr.io/chains-project/breaking-updates:0abf7148300f40a1da0538ab060552bca4a2f1d8-breaking",
  "javaVersionUsedForReproduction": "11",
  "failureCategory": "COMPILATION_FAILURE"
}
\end{lstlisting}

A key contribution of the \benchmarkname benchmark is that each breaking dependency update is packaged as a pair of Docker images.
We include the \emph{preCommitReproductionCommand} and the \emph{breakingUpdateReproductionCommand} in the JSON file to facilitate the reproduction of the breaking update. By running the two commands, it is possible to recreate the error introduced by the version update which causes the project build to fail. The \emph{failureCategory} indicates the cause of the breaking update.

\subsection{Implementation}
\label{implementation}

The complete collection, documentation and storage of the \benchmarkname benchmark is fully automated and is publicly available at \bumpurl. The pipeline is implemented in Java and runs on Java 11. We rely on the library \href{https://mvnrepository.com/artifact/org.kohsuke/github-api}{kohsuke} to query GitHub. We use Apache Maven 3.9.2 for compilation and test execution, Docker 23.0.3 to build reproducible images.

\section{Experimental Methodology}
\label{researchQuestions}

In this section, we introduce the research questions that structure our analysis of the \benchmarkname benchmark, as well as the methodology to answer them.

\subsection{Research Questions}

\newcommand\rqReproducibility{How robust is \benchmarkname regarding the consistent reproducibility of breaking updates?\xspace}

\newcommand\rqFailureTypes{What are the different types of failures behind the breaking updates of \benchmarkname?\xspace}

\newcommand\rqCompileErrors{What kind of changes in libraries have led to compilation and test errors in client projects included in \benchmarkname?\xspace}


\begin{enumerate}[label=RQ\arabic*:, ref=RQ\arabic*]

    \item \label{rq: reproducibility}\textbf{\rqReproducibility}

    The core design principles of \benchmarkname are meant to make the breaking updates reproducible. With this first question, we aim at assessing this property on different operating systems and in complete isolation from the network. This assessment will provide empirical evidence about the degree of reproducibility of \benchmarkname's data.

    \item \label{rq:failures} \textbf{\rqFailureTypes}

    We categorize the types of build failures that cause a breaking update. This analysis provides insights into the various effects that breaking changes can have in client projects. 
    By determining the number of breaking updates in each category, we can study the prevalence of failure types that incompatible library version can incur in client projects.
    
    \item \label{rq:compilatioErrors} \textbf{\rqCompileErrors}

    In this RQ, we focus on the different types of syntactical and behavioral errors that occur in client projects as a result of a breaking dependency update. We specifically aim to find the causes of compilation errors and test errors that are observed in \benchmarkname. Understanding these underlying causes will benefit researchers and developers when designing mitigation strategies for breaking changes.  
     
\end{enumerate}

\subsection{Methodology for RQ1 (Reproducibility)}

The goal of this RQ is to evaluate the robustness of reproducing breaking dependency updates in \benchmarkname, on different platforms. We analyze the different steps of the sanity check cases discussed above in \autoref{sec:sanity}. We check if all tests run successfully on both Windows 11 and Linux(Ubuntu 22.04.2 LTS) platforms in the \precomm execution. 
Next, we analyze the logs generated by the execution of the \breakingcomm and we determine the failure type. We check that the failure type is the same over all runs and all environments. This process allows us to guarantee that  \benchmarkname yields the same breakages over different platforms.

\subsection{Methodology for RQ2 (Failure category)}
\label{sec:meth-rq2}
With RQ2, we aim at precisely characterizing the different types of build failures for each breaking update in \benchmarkname.
In order to categorize the failures, we analyze the log file generated when reproducing the breaking update. 
We search for predefined keywords to automatically identify the type of build failure that occurred. We define the set of keywords after manually analyzing the entire sample of log files.
We categorize a breaking update as \emph{Compilation failure} if the project's compilation fails, as indicated by the presence of the keyword "compilation error" in the log file. If compilation is successful but some test cases fail, we label it as a \emph{Test failure}, identified by keywords such as "There are/were test failures" and other test driver messages. Another type of build failure is a \emph{Dependency resolution failure}, which occurs when Maven cannot locate the new version of dependency in the known package repositories. We identify this failure type when the log file contains error messages related to unresolvable dependencies. Additionally, a breaking update may lead to an \emph{Enforcer failure} if the updated dependency violates dependency rules that are defined in the project under consideration. As there are different types of enforcer rule violations, we use multiple keywords related to different plugins such as Maven enforcer plugin, Maven HPI plugin and Maven Checkstyle plugin to identify this failure type. Similarly, we assign the failure category \emph{Dependency lock failure}, if the failure happened during the execution of the l \emph{check} in the plugin \emph{dependency-lock-maven-plugin}.

\begin{table}[bp]
\caption{\benchmarkname descriptive statistics}
\label{tab:bump_metadata}
\centering

\begin{tabular}{p{6cm}r}
\toprule
\benchmarkname Metadata &  Occurrence\\
\midrule

    Number of breaking update candidates &  \nbTotalReproductions \\

    \begin{tabular}{@{}l@{}}Reproducible breaking dependency updates before \\ sanity check
    \end{tabular}
    & \nbBreakingUpdatesBeforeSanityCheckings \\
 
    Final Reproducible breaking dependency updates &  \nbBreakingUpdates \\
    
    \begin{tabular}{@{}l@{}}Number of projects with at least   
    one reproduced \\  breaking updates
    \end{tabular}
    & \nbUniqueProjects \\
    
    Median number of direct dependencies per project &  \nbDirectDependencies \\

    \begin{tabular}{@{}l@{}}Median number of transitive dependencies \\ per project
    \end{tabular}
    & \nbTransitiveDependencies  \\
    
    \begin{tabular}{@{}l@{}}Median value of effective dependency changes per \\breaking update
    \end{tabular} & \nbDependencyChangesPerBDU \\

    Number of reproduction images&  2x\nbBreakingUpdates= \nbBreakingUpdateImages \\
       
   Median \benchmarkname image size in MB &  \nbMedianDockerImages MB \\

\bottomrule
\end{tabular}
\end{table}

\subsection{Methodology for RQ3 (Breakage cause)}
The goal of this RQ is to investigate deeper into the compilation and test failures and identify their underlying causes. We answer this RQ in two parts, one of which focuses on compilation failures and the other on test failures. To investigate the root causes of compilation failures, we follow the process of \autoref{fig:japicmp-results-to-client}.
We first analyze the log file generated during the breaking update build. By parsing the error messages in the log file, we identify the specific lines in the client project where the compilation errors are located. We then use Spoon, \cite{pawlak:hal-01169705}, to create the abstract syntax tree (AST) of the project's source code and identify the exact method invocations and constructor calls in the identified erroneous code lines. Next, we use the static analysis tool \href{https://github.com/siom79/japicmp}{japicmp} to compare the jar files of the two dependency versions involved in the breaking update and extract all the potential breaking changes. By crossing the extracted information from the project under consideration and from japicmp, we are able to determine the root cause of the compilation failure.

\begin{figure}
\centering
\includegraphics[width=0.80\linewidth]{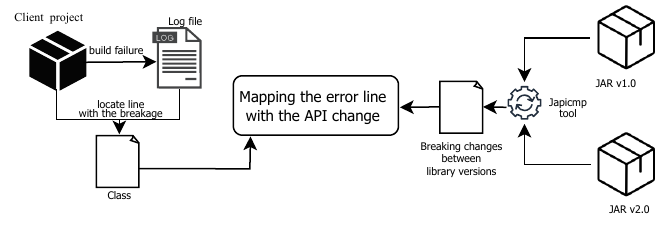}
\caption{Overview of the process to map the API changes with the client project build errors}
\label{fig:japicmp-results-to-client}
\end{figure}

To identify how likely an API change type results in a breakage, we calculate a normalized breakage likelihood metric $\mathcal{S}$ as given in \autoref{eq:normalize-metric}.
\newenvironment{conditions}
  {\par\vspace{\abovedisplayskip}\noindent\begin{tabular}{>{$}l<{$} @{${}={}$} l}}
  {\end{tabular}\par\vspace{\belowdisplayskip}}

\begin{equation} \label{eq:normalize-metric}
 \mathcal{S_t} = \frac{\mathcal{BL_t} - \min(\mathcal{BL_t})}{\max(\mathcal{BL_t}) - \min(\mathcal{BL_t})}
\end{equation}
where:
\begin{conditions}
 t     &  a change type \\
 \mathcal{BL_t}     &  $\frac{N_{breakage}(t)}{N_{all}(t)}$ \\
 N_{breakage}(t)     &  occurrence of $t$  in breakages \\   
 N_{all}(t)          &  occurrence of $t$ in all updates
\end{conditions}

In the second part of the RQ, we investigate the test failures. We collect the test cases that result in errors or failures, as well as the cause of these test failures. The cause can be an exception that interrupts the execution of the test, or it can be an assertion failure. Our analysis is based on the number of occurrences of different exception types, as well as on the number of assertion failures.

\section{Experimental Results}
\label{results}

\subsection{Descriptive Statistics}

\autoref{tab:bump_metadata} summarizes the key metrics of \benchmarkname. We started from \nbTotalReproductions pull-requests on GitHub that are candidate breaking dependency updates. Out of those, we successfully locally reproduced \nbBreakingUpdatesBeforeSanityCheckings.
Next, we discard 47 breaking updates at the sanity check step. The final version of \benchmarkname consists of 571 validated, high-quality breaking updates.
These reproduced breaking dependency updates are performed on \nbUniqueProjects different Java projects. 
\autoref{fig: dependencies} presents the distribution of the number of direct and transitive dependencies in the \nbUniqueProjects projects of \benchmarkname. The median number of direct dependencies per project is \nbDirectDependencies\xspace and the median number of transitive dependencies is \nbTransitiveDependencies. These distributions indicate that \benchmarkname includes a variety of projects, some with few dependencies and others with hundreds of dependencies.
By design, all breaking dependency updates bump a single dependency version. However, per the Maven dependency resolution algorithm, this might ripple in actually updating several sub-dependencies.
This is the case for 44 breaking updates where the updated library is a meta-library (only POM, no Jar) packaging sub-libraries.
Another rippling effect of version bumping is the downstream changes in transitive dependencies. Specifically, for 316 breaking updates in \benchmarkname, there were more than one dependency update effectively applied, triggered the one-line change in the top pom file. A breaking update in \benchmarkname actually has a median value of \nbDependencyChangesPerBDU dependency changes. We observe the maximum number of 81 dependency changes when updating the dependency \texttt{org.jenkins-ci/acceptance-test-harness} in the project \texttt{jenkinsci/code-coverage-api-plugin}. 

\begin{figure}
\centering
\includegraphics[width=0.80\linewidth]{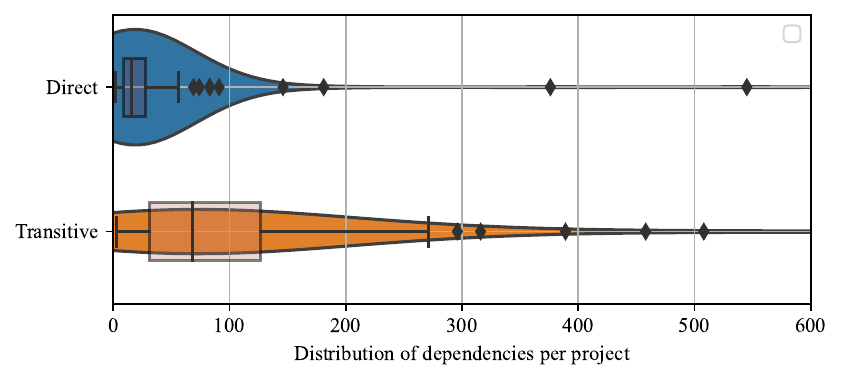}
\caption{Distribution of dependencies per project over \benchmarkname}
\label{fig: dependencies}
\end{figure}

Each successfully reproduced breaking update is packaged in a pair of Docker images for long-term preservation.
\benchmarkname contains a total of \nbBreakingUpdateImages reproducible images.
Each image is created based on a base image with the specifications mentioned in \autoref{implementation}. The images contain a layer with the project before the build and a second layer with the project after the build.
Eighty percent (80\%, 457/\nbBreakingUpdates) of the image pairs have a size between 581 MB and 1 GB. 
The median image pair size value is \nbMedianDockerImages MB.
The smallest \benchmarkname image pair is for a breaking update in \texttt{sabomichal/liquibase-mssql}, with 584 MB. The largest pair of images takes 7.77 GB, it is for a breaking update in  \texttt{google-cloud-java}, which has 621 dependencies. 

\subsection{RQ1 \textbf{\rqReproducibility}}

To address this question, we execute all the breaking dependency updates of \benchmarkname three times, on different platforms. Each breaking dependency update is stored in a pair of images, one \precomm (Pre-breaking image) and one \breakingcomm (Breaking image). For each image, we build the project in a Docker container configured with the \texttt{network none} command to ensure execution without a network connection.

The reproduction of the breaking update is considered successful if 
1) each execution of the image with the \precomm passes without error on all platforms;  and
2) each execution of the image for the \breakingcomm fails due to one single consistent cause. 

Initially, \benchmarkname contains 628 breaking dependency updates as a result of the reproduction process described in \autoref{sec:reproduction-process}. Next, in the sanity check, we discard 42 breaking updates due to flaky tests and 15 breaking updates because the failure causes were related to the system configuration and not to the update itself. As a final result, \benchmarkname consists of \nbBreakingUpdates breaking dependency updates.
We execute the \nbBreakingUpdates images with the \precomm on both platforms. We run the test suite three times in \precomm on \nbUniqueProjects projects to detect flaky tests. We meticulously verify the successful execution of all test cases on both Linux and Windows, by parsing the build log. 

To assess whether 100\% of the failure types on both platforms are consistent with the reference in \benchmarkname, we execute the 571 images containing the \breakingcomm three times. We check that all causes of compilation failures and test failures on Windows and Linux occur due to the same causes as in the reference \benchmarkname run. For example, the update of the dependency \texttt{net.sf.jasperreports/jasperreports} from version \texttt{6.18.1} to \texttt{6.19.1} in the project \texttt{biapi} fails in the compilation execution due to \texttt{incompatible types} in the original reproduction in \benchmarkname.

To measure the complete reproduction of the breaking dependency updates independent of the environment, we execute the breaking updates on both Linux and Windows systems. We discard breaking updates that are not reproducible on both systems. For example, in the project \texttt{alphagov/pay-connector}, the dependency update of \texttt{org.glassfish.jaxb/jaxb-runtime} from \texttt{2.3.5} to \texttt{4.0.0} is successfully reproduced on Linux but fails on Windows due to a test failure. This test failure only occurs on Windows, because in test reverseDnsShouldReturnHostIfIpIsValid, Windows Subsystem for Linux does not support reverse DNS lookup with the default settings, a typical platform dependence problem. Overall, our reproducibility checks allow researchers to experiment with \benchmarkname independently of their experimental platform.

\begin{tcolorbox}[boxrule=1pt,arc=.3em, left=4pt, right=4pt]
  \textbf{Answer to \ref{rq: reproducibility}}: A benchmark of breaking updates is easily unreproducible and unsound due to test flakiness and implicit platform dependence. In \benchmarkname, we take special care of mitigating those risks. \benchmarkname contains \nbBreakingUpdates breaking dependencies updates that have been validated three times, on two different platforms (Linux and Windows). For 100\% of these executions the reproduction is successful: all pre-breaking commits build correctly and all \breakingcomm failures are consistently the same.
  \benchmarkname provides fully reproducible breaking updates that will be used by future research on the important research topic of software dependency engineering.
\end{tcolorbox}

\subsection{RQ2 \textbf{\rqFailureTypes}}

With this RQ, we aim to categorize the different build failures related to dependency updates in \benchmarkname. In  \autoref{tab:failure-category-counts}, we summarize the 5 main categories of failures triggered by dependency updates:  
compilation failure is a failure in the compilation phase; test failures are identified during test execution; enforcer failures group errors triggered by the dependency usage rules; dependency lock failures are identified before the compilation process begins; dependency resolution failures are related to missing artifacts, see \autoref{sec:meth-rq2}.
In \autoref{tab:failure-category-counts}, we report the number of occurrences of each failure category in \benchmarkname. 

Out of the \nbBreakingUpdates breaking updates in \benchmarkname, \nbCompilationBreakage (\percentageCompilationBreakage\%) are due to compilation failures and \nbTestingBreakage (\percentageTestBreakage\%) are due to test failures. 
In these cases, the new version of the dependency introduces syntactic or behavioral changes in the API that cause failures in the build process of the clients.

We look into the details of the breaking dependency updates of the \texttt{IDS-Messaging-Services} project. \benchmarkname includes 35 breaking dependency updates categorized as compilation failures, and 21 categorized as test failures for this project. 
For instance, when the developers of \texttt{IDS-Messaging-Services} updated \texttt{org.springframework/spring-webmvc} from version \texttt{5.3.24} to version \texttt{6.0.5}, the CI failed because of a compilation failure related to missing symbols. Another update of the \texttt{com.fasterxml.jackson.core/jackson-databind} dependency from version \texttt{2.9.10.8} to version \texttt{2.13.3} triggered 6 test failures in the sub module \texttt{messaging}.
In \autoref{RQ3} we provide an extensive analysis of the causes that lead to these compilation and test failures. 

Enforcer failures represent \percentageEnforcerBreakage\% (\nbMavenEnforcerBreakage) of the breakages in  \benchmarkname. 
Enforcer rules are additional rules, defined outside the default Maven build process, which encode interoperability constraints between libraries. Enforcer failures are due to the violation of at least one of these rules.
For example, the project \texttt{pac4j/dropwizard-pac4j} defines an enforcer rule on dependency convergence. This rule checks whether the indirect dependency versions converge, i.e. if there are two dependencies A and B declared in a project, and both dependencies depend on another dependency C, then both A and B should depend on the same version of C, otherwise the rule will fail. When the project \texttt{pac4j/dropwizard-pac4j} updates \texttt{org.eclipse.jetty/jetty-server} dependency, from version \texttt{9.4.35.v20201120} to version \texttt{9.4.46.v20220331}, the dependency convergence rule is violated. 
It is violated because both the dependencies \texttt{io.dropwizard/dropwizard-testing} and \texttt{org.eclipse.jetty/jetty-server} depend on the dependency \texttt{org.eclipse.jetty/jetty}, but after updating \texttt{org.eclipse.jetty/jetty-server}, they depend on different versions of \texttt{org.eclipse.jetty/jetty}. 
These enforcer failures highlight the fact that, occasionally breaking updates are not only intrinsically due to breaking changes in APIs, but are also triggered by specific client project constraints. Therefore, an area of improvement for dependency management bots would be to take these project constraints into account when the dependency updates are performed.

\begin{table}[bp]
\caption{The number of breaking updates per failure category in the \benchmarkname benchmark}
\centering

\begin{tabular}{lc}
\toprule
Failure category  &  \makecell{Number of breaking updates} \\ 
\midrule
  Compilation failure &  \nbCompilationBreakage (\percentageCompilationBreakage\%) \\
  Test failure &  \nbTestingBreakage (\percentageTestBreakage\%) \\
  Enforcer failure &  \nbMavenEnforcerBreakage (\percentageEnforcerBreakage\%)  \\
  Dependency lock failure &  \nbDependencyLockBreakage (\percentageDependencyLockFailure\%)  \\
  Dependency resolution failure &  \nbDependencyResolutionBreakage (\percentageDependencyResolutionFailure\%) \\ 
 \bottomrule
 
\end{tabular}
\label{tab:failure-category-counts}
\end{table}

\begin{table}[bp]
\caption{The 10 most common causes for compilation failures in the \benchmarkname benchmark}
\centering
\label{tab:compilaton-error-counts}

\begin{tabular}{lr}
\toprule
Causes  &  \makecell{Total number \\ of occurrences}\\
\midrule
  cannot find symbol &  \nbCannotFindSymbol \\

  cannot access &  \nbCannotAccess \\

  package <> does not exist &  \nbPackageNotExist \\

  \begin{tabular}{@{}l@{}}constructor <> in class cannot be  \\ applied to given types\end{tabular}
  &  \nbConstructor  \\

  \begin{tabular}{@{}l@{}}method does not override \\ or implement a method from a supertype \end{tabular}
  &  \nbMethOverride  \\
 
  static import only from classes and interfaces &  \nbStaticImport  \\ 

  incompatible types: <> cannot be converted to <> &  \nbIncompatibleTypes \\

  reference to <> is ambiguous &  \nbReferencesAmbiguous  \\

  method <> cannot be applied to given types &  \nbMethodTypes  \\

  no suitable constructor found for <> & 5 \\
\bottomrule

\end{tabular}
\end{table}

\benchmarkname includes 14 dependency update breakages caused by lock failures. This type of failure occurs when the project uses the \href{https://github.com/vandmo/dependency-lock-maven-plugin}{\texttt{dependency-lock-maven-plugin}} to lock versions of direct and transitive dependencies. In this particular case, none of the project's dependencies can be successfully updated by individual pull requests with single-line version bumps. The presence of this type of failure stresses the need for support of dependency lock plugins by dependency management bots.   

The least common type of failure in \benchmarkname is due to dependency resolution errors. This occurs when the updated version of the dependency or a transitive dependency of the new version cannot be resolved by Maven. We observe this failure in \nbDependencyResolutionBreakage (\percentageDependencyResolutionFailure\%) breaking dependency updates in \benchmarkname. For example, in the project \texttt{google-cloud-java}, the build process fails when updating the \texttt{google-cloud-shared-dependencies} dependency from version 3.1.0 to version 3.1.1-SNAPSHOT. The build failure happens because the dependency 3.1.1-SNAPSHOT does not exist in the Maven central repository. To solve this issue, the developers of \texttt{google-cloud-java} eventually added the dependency to the repository as a part of the project itself, after the failed version update attempt.

\begin{tcolorbox}[boxrule=1pt,arc=.3em, left=4pt, right=4pt]
  \textbf{Answer to \ref{rq:failures}}: 
  \benchmarkname contains \nbBreakingUpdates breaking dependency updates, \percentageCompilationTestBreakage\% correspond to classical breaking dependency updates due to compilation or test failures.
  While those problems are known, it is the first time that they are encapsulated in a fully reproducible manner. 
  The equal proportion of compilation and test failures suggests an equally important need to investigate automatic migration of APIs to address compilation errors and automatic repair of behavioral changes in dependencies. 
  \benchmarkname also includes \percentageAllOtherBreakage\% of more original failures that were never discussed before in the literature: enforcer and locking failures. The \benchmarkname benchmark contains real-world examples of breakages for each failure type, it will serve as foundation for future research on all facets of the breaking update problem.
\end{tcolorbox}

\subsection{RQ3: \textbf{\rqCompileErrors}}
\label{RQ3}

\begin{table}[htbp]
\caption{The 10 most common API changes that caused compilation errors in client projects included in \benchmarkname}
\centering
\label{tab:compilaton-error-causes}

\begin{tabular}{lrr}
\toprule
API change & \makecell{Total \\ Count} & \makecell{Normalized breakage \\ likelihood score ($\mathcal{S}$)} \\
\midrule

  method removed & 139 & 0.42\\

  class removed & 30 & 1.00\\

  constructor removed & 25 & 0.36\\ 
 
  class generic template changed & 7 & 0.13\\

  method now throws checked exception & 2 & 0.39\\
  
  method removed in super class & 2 & 0.06\\

  method return type changed & 2 & 0.04\\

  method abstract added to class & 1 & 0.55\\

  \begin{tabular}{@{}l@{}}method no longer \\ throws checked exception
  \end{tabular} & 1 & 0.24\\

  method parameter generics changed & 1 & 0.00\\

  no change detected by japicmp & 130 &  N/A \\

\bottomrule
\end{tabular}
\end{table}

Now, we investigate the 431 \benchmarkname (243+188) breaking updates triggered by either compilation or test failures. In the first part of the RQ, we categorize the causes for compilation errors and link them to API changes in the updated dependency that caused the errors. \autoref{tab:compilaton-error-counts} presents the 10 most common causes of compilation errors that we observe in the compilation logs. The total count of errors is higher than the number of breaking updates since one build failure may be caused by more than one error. For instance, when \texttt{NemProject} updated \texttt{org.flywaydb/flyway-core} from version \texttt{3.2.1} to \texttt{9.21.1}, resulted in 4 missing symbol errors and 1 constructor incompatibility error.

The error \emph{cannot find symbol} occurs whenever the Java compiler cannot recognize an identifier. It is the most common cause for compilation failures. When a used class, a method, or a constructor is removed in the updated dependency, the client project cannot build. The second most common error \emph{cannot access} occurs when a class/method/field visibility is changed in the updated dependency, for example from `public' to `private'. The error messages \emph{package <> does not exist} and \emph{static import only from classes and interfaces} are related to the two errors mentioned above. Among the other common errors, \emph{constructor <> in class cannot be applied to given types} and \emph{no suitable constructor found for <>} are directly related to constructor signature changes, and the errors \emph{method does not override or implement a method from a supertype} and \emph{method <> cannot be applied to given types} are associated with method signature changes. The overall proportions of error occurrences we observe are comparable to the results of Jayasuriya \etal\cite{jayasuriya2023}. Even though the order of categories slightly differs, out of the top 10 compilation errors reported in their study, 8 also appear as the most common compilation errors in our study. 

Next, we retrieve the changes in the API of libraries that lead to breaking updates. We use the japicmp tool to get a list of changes between the two versions of an API involved in a breakage. Then, we map the changes to the error observed in the client project. This mapping allows us to retrieve the relevant API changes that are accountable for causing compilation errors in client projects. In \autoref{tab:compilaton-error-causes} we provide the total count of occurrences for the 10 most breakage-prone API changes and corresponding breakage likelihood scores. The majority of the compilation errors observed in \benchmarkname are indeed caused by the removal of a class, a method, or a constructor. This is fully consistent with \autoref{tab:compilaton-error-counts} with the most common compilation error, \emph{cannot find symbol}. Moreover, when comparing with the results obtained in similar studies done by Ochoa \etal\cite{ochoa2022breaking} and Jayasuriya \etal\cite{jayasuriya2023}, we observe that these three removals and method return type alterations appear in the top 10 changes in all three studies. This consistency suggests that these four types of API changes are universally prone to introducing breakages in client projects. On the other hand, for the API changes \emph{method abstract added to class} and \emph{method now throws checked exception}, the likelihood scores indicate that they are clear causes for breakages, despite their rare occurrence. We further notice that some compilation errors in client projects are caused by updated transitive dependencies but not by the updated dependency itself. For this reason, for 130 cases, we could not find any related API changes using japicmp. 

In the second part of the RQ, we investigate the test failures observed in \benchmarkname.  We distinguish between test execution errors, which occur due to a crash during the execution of a test case, and test assertion failures, which occur when the project under test does not behave as expected by a test assertion. In \autoref{tab:test-failures}, we present the number of breaking updates and the total number of test cases in those projects that fail due to different types of test execution errors and test assertion errors. All rows in the table except for the last row contain the different types of execution errors we observe, and the last row presents the results for the assertion errors. It is evident from the results that more than half of the test failures that trigger a breaking update are caused by test execution errors. 
\begin{table}[tbp]
\caption{Causes of test failures. Some breaking updates have more than one type of error.}
\centering
\label{tab:test-failures}
\begin{tabular}{lrr}
\toprule
Cause of test failure & \makecell{Number of \\ test cases} & \makecell{Number of \\ breaking updates} \\
\midrule
  NoClassDefFoundError  &  1751  &  68 \\
  IllegalStateException  &  1500  &  13 \\
  ClassCastException   &  386  &  28 \\
  UnsupportedClassVersionError  &  204  &  11 \\
  Other exceptions &  434 &  83 \\
  Test assertion errors &  263 &  60 \\
\bottomrule
\end{tabular}
\end{table}

The most common error \emph{NoClassDefFoundError} occurs due to an incomplete runtime classpath, which is a consequence of a disappearing class in the updated dependency.
\emph{IllegalStateException} are due to changes in the contract of API methods that are detected by an application assertion.
\emph{UnsupportedClassVersionError} occurs when the updated dependency or one of its transitive dependencies has been compiled using an incompatible Java version. The rest of the error types mostly occur due to project-specific functionalities. For example, in the project \texttt{jadler/jadler-mocking}, the dependency update of \texttt{org.slf4j/slf4j-api} from version \texttt{1.7.36} to \texttt{2.0.0} has caused 5 \emph{classCastExceptions} in 5 test cases out of 235 total test cases. The underlying cause for the \emph{classCastException} in this case is the class \texttt{net.jadler.JadlerMockerTest. addAppenderToStream} not obtaining the expected \texttt{ch.qos.logback.classic.LoggerContext}, but instead getting \texttt{org.slf4j.helpers. NOPLoggerFactory} because an appropriate \texttt{slf4j} binding was not found on the classpath. To the best of our knowledge, this is the first study that reports on and analyzes in detail test execution errors in breaking updates.
Meanwhile, test assertion errors only contribute to 60 failures of the 188 total test failures.
These observations reveal that the breaking updates are more often related to clear erroneous runtime states that throw exceptions, and more rarely due to subtle behavior changes in the behavior of the API. We believe this is good news as frank breakages are easier to detect, understand and resolve.

\begin{tcolorbox}[boxrule=1pt,arc=.3em, left=4pt, right=4pt]
  \textbf{Answer to \ref{rq:compilatioErrors}}: Our analysis of compilation failures shows that \benchmarkname captures a wide variety of API changes in libraries causing breaking updates incl. class and method removal. 
  Interestingly, we are the first to show that the breakages caught at runtime, during test execution, are mostly exceptions with an explicit cause, and more rarely due to subtle behavior changes captured by a test assertion.
  This demonstrates that developers of Java libraries have a healthy culture of API contract checking with fail-fast assertions. 
\end{tcolorbox}

\section{Related Work}
\label{relatedWorks}

Engineering updates of third-party dependencies is a pressing challenge in software development, which has received research attention.

Hejderup \etal \cite{HEJDERUP2022111097} assess the effectiveness of test suites in identifying compatibility issues arising from dependency updates of Java projects. The study examines 521 projects, evaluates the coverage and effectiveness of test suites, and highlights the need for static analysis to address coverage limitations.

Empirical studies have shed light on the impact of breaking changes within software ecosystems.
Decan \etal \cite{Decan2017} explore the challenges related to package dependency updates. The authors observe that a majority of packages declare dependencies, and the proportion of such packages increases over time.
Ventury \etal\cite{venturini2023} investigate the impact of breaking changes on 384 npm client packages, finding that modifications to functions, changes in dependencies, and alterations in data types are key contributors to breaking versions.
Keshani et al.\cite{Keshani2023} identify a breaking change as an instance where the method signature is removed or altered within a major release. 

Along the same lines, previous research has analyzed the impact of breaking changes on clients. 
Xavier \etal\cite{xavier2017historical} presents a large-scale study on API breaking changes in Java libraries. The study analyzed 317 real-world Java libraries, 9K releases, and 260K client applications. The results show that 14.78\% of API changes break compatibility with previous versions and 2.54\% of clients are impacted.
Raemaekers \etal\cite{raemaekers2016} employed the tool called Clirr \cite{clirr2012} to analyze two jar files and identify the changes in public APIs. Their findings reveal that breaking changes are commonly found even in non-major releases and that they cause a significant impact on clients with compilation errors.
Ochoa \etal\cite{ochoa2022breaking} improve the study by Reamaekers \etal\cite{raemaekers2016} by expanding the data by using a bigger dataset, and by using an improved Java bytecode analysis tool called Maracas to identify breaking changes between two libraries.
Macho \etal\cite{macho2018automatically} presents an approach called BUILDMEDIC to automatically fix dependency-related compilation errors in Maven build files. The authors extracted changes to Maven build specifications from 23 open-source Java projects and analyzed 37 revisions to identify the specific changes that fixed the dependency-related compilation breakage.
Jayasuriya \etal\cite{jayasuriya2023} also presents an empirical study on how the breaking dependency updates affect the client projects in the Maven ecosystem. The results obtained from their study show that the majority of compilation errors in client projects occur due to changes in transitive dependencies. While our paper also analyzes the causes of breakage, none of those related works have engineered a benchmark of breaking updates.

In recent research, different software bug benchmarks have been built for software testing and automatic program repairing. Defects4J \cite{Gay2020} contains 835 reproducible bugs collected from 17 projects, and Bugs.jar \cite{Saha2018} contains 1,158 reproducible bugs collected from eight Apache projects.
Kabadi \etal\cite{kabadi2023} constructs a benchmark consisting of 102 bugs that were detected through continuous integration failures in 40 substantial real-world programs.
The Java bug benchmark BEARS \cite{madeiral2019bears} contains 251 bugs from 72 GitHub projects, where each bug and its patch are identified through Travis-CI \cite{Beller2017}. 
When considering the bug repositories created for C language, IntroClass focuses on small programs written by beginners and contains 998 bugs collected from programs written by students, whereas ManyBugs contains 185 bugs collected from 9 open source programs \cite{LeGoues2015}. QuixBugs \cite{Lin2017} is a multi-language benchmark containing bugs from 40 programs implemented in both Python and Java.

A noteworthy software bug analysis technique is BugSwarm, as described in Tomassi et al.'s work \cite{tomassi2019bugswarm}. BugSwarm stands out for its ability to automatically extract software defects from GitHub-hosted projects and encapsulate them within reproducible containers.
BugBuilder \cite{jiang2022bugbuilder} identifies only the most relevant bug fix patches while excluding unrelated code changes in the patch commits.

In contrast to these benchmarks, \benchmarkname focuses on bugs related to breaking dependency updates. 
\benchmarkname places a significant emphasis on ensuring reproducibility by encapsulating each breaking dependency update within Docker images. This approach guarantees the long-term reproducibility of each breakage for sake of future research.

The most closely related work to our study is CompSuite \cite{Xu2023CAD}. CompSuite provides a dataset of real-world Java client-library pairs where upgrading the library causes compatibility issues in the corresponding clients. It includes 123 incompatible client-library pairs, each associated with a test case that can be used to reproduce the issue.
There are three main differences with our study compared to CompSuite. \benchmarkname consists of breaking updates attempted by developers in the real-world, while in CompSuite the breaking changes are artificial and seeded by the authors themselves to detect incompatible versions. 
Second, CompSuite only focuses on test case failures incurred by dependency updates, but our study analyses multiple types of build failures as mentioned in \autoref{tab:failure-category-counts}. 
The last difference is that CompSuite does not guarantee reproducibility, as it does not provide a mechanism to isolate all the artifacts involved in the build. \benchmarkname provides an isolation mechanism for breaking updates through Docker images that ensures the reproducibility of each breaking dependency update.

\section{Conclusion}
\label{conclusion}

In this paper, we have introduced \benchmarkname, a benchmark of \nbBreakingUpdates breaking dependency updates collected from \nbUniqueProjects Java projects on GitHub. \benchmarkname stores the breaking updates as Docker images, which can be executed without network connectivity, guaranteeing reproducibility of the breakages over the long-term.
We ensure the full reproducibility of \benchmarkname on two notable platforms, specifically Windows 11 and Linux/Ubuntu 22.04.2 LTS.  

To our knowledge, \benchmarkname is the first benchmark of real-world, fully reproducible breaking dependency updates. 
The benchmark is publicly available at \bumpurl.
We believe that \benchmarkname is a sound foundation for reproducible research on dependency updates and invite the research community to use it via the public repository. 

\section{Acknowledgments}

This work was supported by the CHAINS project funded Swedish Foundation for Strategic Research (SSF), the WebInspector project funded by Swedish Research Council (VR), as well as by the Wallenberg Autonomous Systems and Software Program (WASP) funded by the Knut and Alice Wallenberg Foundation.

\IEEEtriggeratref{25}
\bibliographystyle{IEEEtran}
\bibliography{main}

\end{document}